# Astro2020 Science White Paper

# The Chemical/Dynamical Evolution of the Galactic Bulge

**Thematic Areas:**

☐Stars and Stellar Evolution

☐Resolved Stellar Populations and their Environments

☐Galaxy Evolution

**Principal Author:**
Name:	R. Michael Rich
Institution:  Department of Physics and Astronomy, UCLA
Email: rmr@astro.ucla.edu
Phone: 310-775-5138

**Co-authors:** (names and institutions)

Adam Burgasser, UC San Diego, aburgasser@ucsd.edu
Will Clarkson, University of Michigan, Dearborn
Christian Johnson, Harvard/Smithsonian CfA,
Andrea Kunder, St. Martin's University

**Abstract**:  The last decade has seen apparent dramatic progress in large spectroscopic datasets aimed at the study of the Galactic bulge.  We consider remaining problems that appear to be intractable with the existing data, including important issues such as whether the bulge and thick disk actually show distinct chemistry, and apparent dramatic changes in morphology at Solar metallicity, as well as large scale study of the heavy elements (including r-process) in the bulge.  Although infrared spectroscopy is powerful, the lack of heavy element atomic transitions in the infrared renders impossible any survey of heavy elements from such data.  We argue that uniform, high S/N, high resolution data in the optical offer an outstanding opportunity to resolve these problems and explore other populations in the bulge, like RR Lyrae and hot HB stars.

**A. Introduction & Context:** Galaxy bulges constitute major components of the visible light from disk galaxies, and are important actors in galactic evolution (e.g. Rich 2013, McWilliam 2016, Barbuy, Chiappini & Gerhard 2018). The Galactic bulge should be the calibrator object for galaxy bulges throughout the Universe, but the bulge is a challenging observational target due to the high extinction, extreme crowding, superposition of populations, and an intrinsically highly complex environment.

Since Astro 2010, observations have conclusively shown that the majority of mass in the bulge is in the bar (e.g. Shen et al. 2010), but many problems in the formation and chemical evolution of the bulge remain unsolved and debated. One fundamental fact remains: the Milky Way bulge is 100 times closer than the nearest comparable example, the bulge of the great galaxy of Andromeda (M31), and we can measure distances, proper motions, radial velocities, and compositions for individual stars in the Milky Way as faint as 20 mag, depending on the band, something that will not be possible in the next decade for M31. Here we argue for a robust *spectroscopic effort at high S/N and spectroscopic resolution* to break many of the key observational and theoretical degeneracies that still remain. The optical is required because it offers the only opportunity to work with the heavy elements and has access to more elements.

What value does Galactic archeology have in an era when the entire star formation history of the Universe from z=9 to the present can be observed? As is frequently stated, one cannot connect the observation of any individual high-redshift galaxy with its progeny, while in principle, the formation and evolution of the Milky Way is memorialized in its fossil record. Unfortunately, the most difficult part of that record to measure and to observe is that of chemical composition; the measurements are difficult and expensive, and the interpretation (chemical evolution models) remain on a far less firm footing than models of dynamics.

**B. Current difficulties and controversies:** Significant advances have been made in the acquisition of large datasets, many of which remain under analysis. APOGEE, the GAIA-ESO Survey, and Galah have all recently completed, and SDSS V will launch in a few years. There is no shortage of spectroscopy, either in archives or on planned; matching with Gaia can provide proper motions, but not parallaxes, as the bulge is too distant. Nonetheless, it is fair to ask what significant questions remain unanswered. We identify two difficulties in particular:

**1. Non-unique interpretation of large datasets:** Perhaps the most important current difficulty is the interpretation of the vast floods of spectroscopic data. Figure 1 below presents one such conundrum. Zoccali et al. (2017) argue that there is strong evidence that the spatial distribution of the bulge changes dramatically from spheroidal (for stars with [Fe/H]<0) to disk-shaped for stars with [Fe/H]>0, yet no strong kinematic break at Solar metallicity is evident (the kinematic signature of the bar sets in at [Fe/H]=-0.5 and does not change). Are we observing direct evidence of the dissipative collapse of the bulge? However, the RR Lyrae (see Fig 5 below) exhibit a more round spatial distribution, but at metallicity 1-2 orders of magnitude lower. How can both findings be consistent? Perhaps the answer is that the more metal rich stars are younger, hence their concentration in a disk?

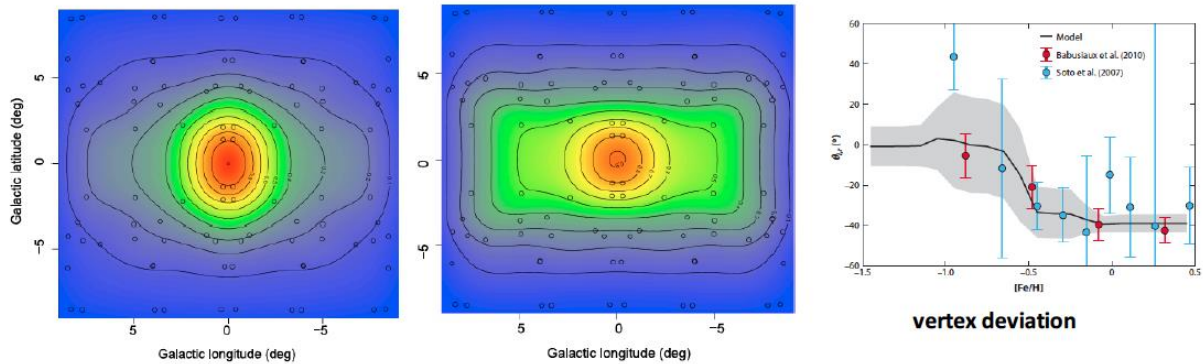

**Fig 1-** Spectroscopy using the Ca infrared triplet finds evidence for a sharp break in spatial properties for the bulge at [Fe/H]=0; the more metal rich stars are in a disk structure; such a dramatic change in spatial structure at Solar metallicity is unexpected (Zoccali et al. 2017). Combining proper motion in galactic latitude and radial velocity gives the *vertex deviation*, which shows no change at Solar metallicity (Barbuy, Chiappini, & Gerhard 2018); this would appear to be inconsistent with the result show in the left-hand panel.

**2. Apparently contradictory observational evidence for the age distribution:** Figure 2 shows one of the enduring controversies from the Galactic bulge, still very much alive and problematic in 2019. The color-magnitude diagram from Clarkson et al. (2008) is derived by applying a proper motion veto of foreground disk stars in the deep HST "SWEEPS" field and is consistent with a ~10 Gyr bulge with little young component (Clarkson et al. 2008). An analysis of four proper motion-cleaned HST fields from the WFC3 Bulge Treasury Survey by Renzini et al. (2018) argues from comparison of the luminosity functions that both metal rich and metal poor populations are ~10 Gyr in age, and likely not younger than 7 Gyr.

Spectra of dwarfs whose light has been amplified by a microlensing event appear to tell a vastly different story (Bensby et al. 2017). Using a self-consistent high resolution abundance analysis, in which [Fe/H], Teff, and *log g* are derived from the stellar spectrum, spectroscopic "parallaxes" are derived, and the implied ages are extremely young. We have a serious contradiction that will require high resolution imagery that spans the entire Galactic bulge, with proper motion data and spectroscopy, to solve.

**C. The need for new spectroscopic datasets:** While there has been an acknowledged huge increase in the number of high resolution spectra, the spectra are not actually what are needed the most. Much of the GES dataset was lower S/N or missed some of the redder spectral regions optimal for bulge studies, while Galah is running on a 4m telescope (as will 4MOST; R=18,000) with limited S/N, spectral coverage, and resolution, and is not observing most of the bulge. Figure 3 shows detail near lines of Eu II and other r-process elements (Sneden et al. 2008). The vast trove of data from APOGEE (and SDSS-V) cannot constrain [La/Eu], or other heavy elements, in the bulge because no transitions are available in the H-band. Medium resolution spectra are actually inadequate for the problem. The lowest acceptable resolution is 20,000 (see Johnson et al. 2014) and the highest possible resolution (e.g. 45,000) would be better; such is planned by the Maunakea Spectroscopic Explorer.

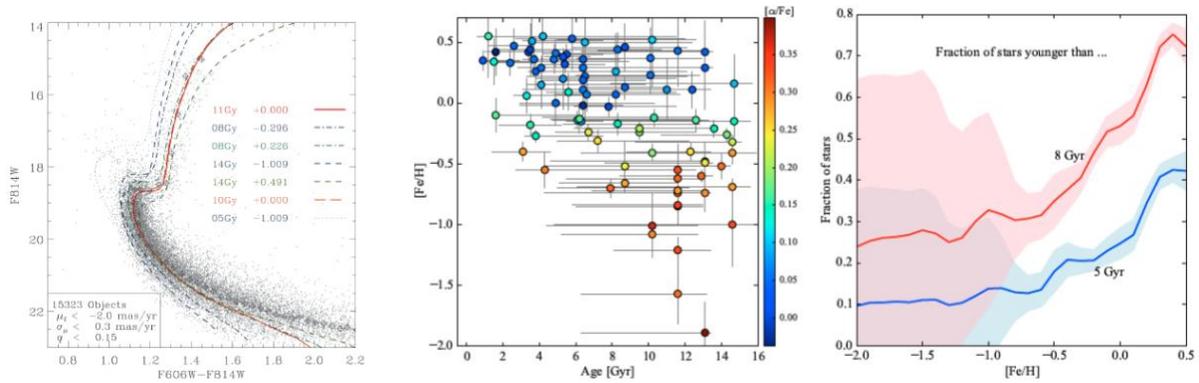

Fig 2- (Left) color-magnitude diagram based on HST ACS data showing the bulge dominated by an old population (Clarkson et al. 2008). (Right panels) Spectroscopic analysis of dwarf spectra obtained during a bulge microlensing event yield log g, Teff, and [Fe/H]. The stars are placed on isochrones allowing an age to be derived from the spectrum. The results are consistent with almost all bulge sitars with [Fe/H]<-0.5 being intermediate age; this would contradict the CMD at left and a new HST-based analysis (Renzini et al. 2018).

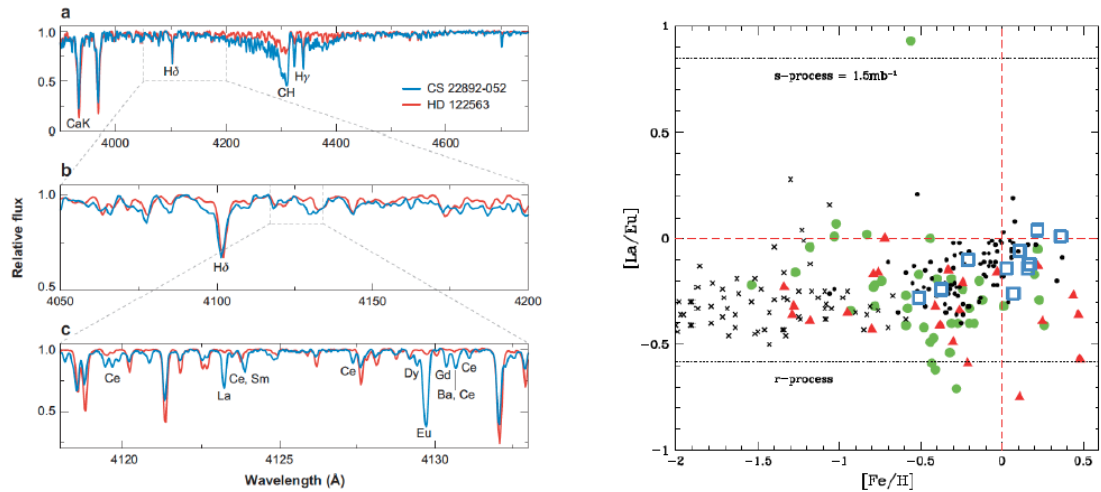

Fig 3- (Left panel) Sneden et al. (2008) shows the importance of high spectral resolution in exploring the heavy elements. The upper spectrum is at R=2500, roughly SDSS 1 resolution. Detail is expanded near 4050A in the second panel. The lower panel shows a spectrum at R=40,000, which reveals many lines of r-process elements in the r-enhanced stars CS22892-052. (Right panel) [La/Eu] (s/r ratio; a proxy for intermediate age to old-  vs [Fe/H] from McWilliam (2016); the black symbols are thick disk stars, while colored symbols  are measurements in the bulge from various studies. With the small samples plotted here, there is the suggestion that the bulge may be enhanced in the r-process, perhaps consistent with rapid, early formation history. Significantly larger, uniform samples obtained at high spectral resolution are needed.

Figure 3 shows the state of the art for surveys of the r/s ratio on the bulge; Johnson, McWilliam & Rich (2013) discovered one extremely enhanced r-process star like those found in the halo, that is likely a bulge member.    The La/Eu ratio gives one a completely different handle on the star formation history, as McWilliam (2016) discusses.  The s-process is produced in the envelopes of intermediate mass AGB stars, while the r-process is produced in massive star SNe, either directly or via neutron star mergers. If we had more data with higher precision, we might well be able to discern a clear difference between the bulge and thick disk [La/Eu] in Fig 3.

In Figure 4, we turn to the light elements whose enhancements in the bulge were first found by McWilliam & Rich 1994. We consider results from APOGEE (Zasowski et 2018), a compilation of oxygen measurements from Barbuy et al. (2018), and K-band CRIRES (VLT) R=50,000 analysis from Nandakumar et al. (2018). All of these studies are based on high resolution spectroscopy, yet all are in serious disagreement, especially at the metal rich end. It is difficult to state with confidence that the bulge and thick disk are different. Indeed, it can be said that the big questions we would like to answer cannot really be addressed with these datasets.

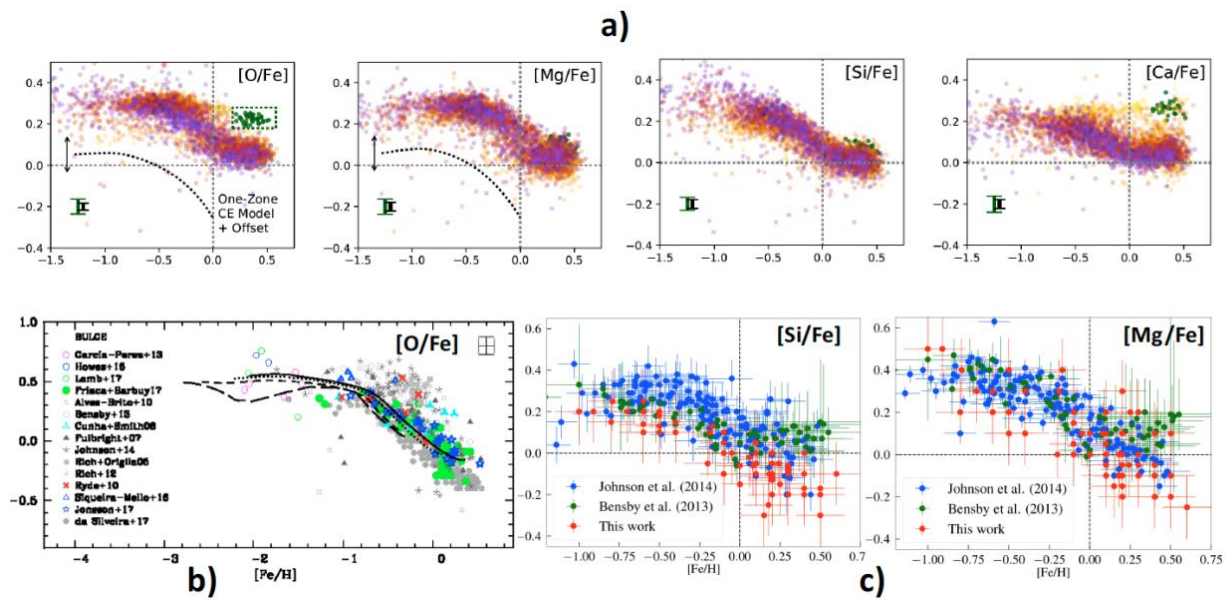

Fig 4-Large optical and infrared spectroscopic surveys do not agree. a) Light elements from APOGEE; Zasowski et al. (2018); b) Oxygen from a range of studies; Barbuy et al. (2018); c) Nandakumar et al. (2018) CRIRES (R=50,000) spectra for giants with |b|<2 deg, including data from optical studies by Bensby and Johnson. *Notice the disagreement at the metal rich end; uncertainties totally blur any distinction between bulge and thick disk, and offer no reasonable constraint on chemical evolution models.*

**D. Successful automated abundance analysis is a long ways off.** Semi-automated analysis of vast datasets won't soon improve matters, as shown in Figure 5. Johnson et al. (2014) reanalyzed an archival VLT FLAMES dataset. Although the older analysis did not use a state of the art method, the problems that occurred at the metal rich end still happen today. Cool, metal rich, stars are full of molecular lines such as CN and TiO, in addition to the vast numbers of atomic lines from mostly iron peak elements. In its initial application, a method called the "Cannon" (Ness et al. 2015) offered the rapid, automated analysis of ~100,000 spectra; however, when forced to derive e.g. Ca abundances using only Ca lines, the 0.6 dex dispersion nullified the results.

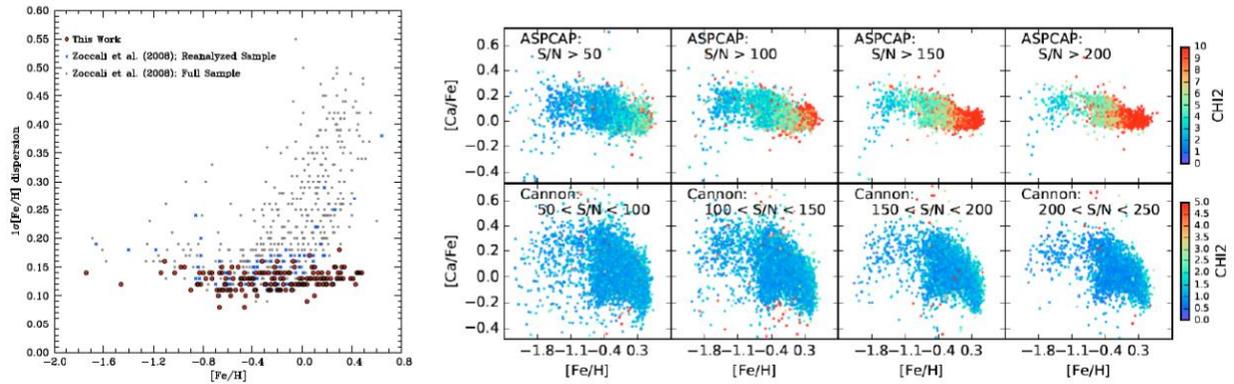

**Fig 5- Two cases where automated abundance derivations are unsuccessful. Left panel is from Johnson et al. 2014. Use of the daospec routine in Zoccali et al. 2008 gives very large errors in abundances, up to 0.5 dex. (Right panel) When the Cannon (Ness et al. 2015) is required to derive abundances only from actual lines of Ca, rather than the entire spectrum, the errors in the derived Ca abundance skyrocket to 0.6 dex (Holtzman et al. 2018).**

Figure 6 shows an exciting stellar population, probably the oldest bulge population, the RR Lyrae stars. Like the blue HB stars that are found all over the bulge, RR Lyrae have Teff mostly hotter than 5000K, making them too hot (and faint) for useful large scale infrared spectroscopic abundance studies like APOGEE or SDSS V. These ancient, metal poor populations might have some of the most interesting objects with unusual heavy element abundance patterns, but are rare so a very significant survey is required to find them (Howes et al. 2016, Koch et al. 2016; Johnson et al. 2013).

The highest resolution spectroscopy, covering the widest wavelength range, remains expensive. TACs are unlikely to allocate a full night on the VLT to obtain 8 or 16 more spectra with UVES in the "giraffe" mode; a dedicated facility to obtain thousands of spectra at R=50,000 and high S/N, like the Maunakea spectroscopic Explorer would be best equipped to solve many remaining problems in the formation of the bulge.

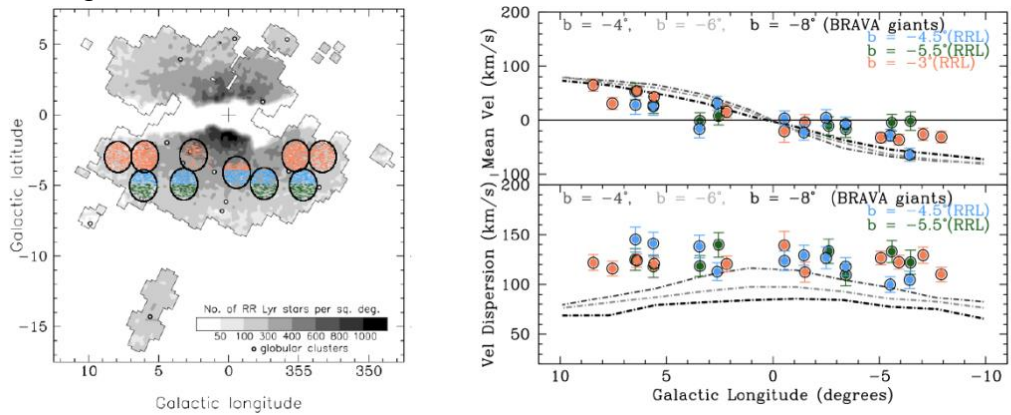

**Fig 6-RR Lyrae in the bulge are a possible new stellar population, which have slower rotation and higher velocity dispersion than stars in the bar. Spatial distribution shown at left (Kunder et al. 2019 in prep.) We would like to obtain high S/N, high dispersion spectra for these stars, which appear also to exhibit kinematic substructure. Possible origins include an ancient merger event, early burst in the history of the Milky Way, or a first phase in the formation of the Galactic bulge. The era may have included r-process enhanced stars.**